# Ultrafast valleytronic logic operations

Francesco Gucci[1], Eduardo B. Molinero[2], Mattia Russo[1], Pablo San-Jose[2], Franco V. A. Camargo[3], Margherita Maiuri[1], Misha Ivanov[4,5,6]*, Álvaro Jiménez-Galán[2,4], Rui E. F. Silva[2,4], Stefano Dal Conte[1], Giulio Cerullo[1,3]*

[1]Department of Physics, Politecnico di Milano, Piazza Leonardo da Vinci 32, Milano 20133, Italy.

[2]Instituto de Ciencia de Materiales de Madrid (ICMM), Consejo Superior de Investigaciones Científicas (CSIC), Sor Juana Inés de la Cruz 3, 28049 Madrid, Spain.

[3]CNR-IFN, Piazza Leonardo da Vinci 32, Milano 20133, Italy.

[4]Max-Born-Institute, Max-Born Strasse 2A, D-12489, Berlin, Germany.

[5]Department of Physics, Humboldt Universität zu Berlin, Berlin, Germany.

[6]Technion—Israel Institute of Technology, Haifa, Israel.

*Corresponding authors. Email: giulio.cerullo@polimi.it, mikhail.ivanov@mbi-berlin.de

## Abstract

Information processing currently reaches speeds as high as 800 GHz. However, the underlying transistor technology is quickly approaching its fundamental limits and further progress requires a disruptive approach. One such path is to manipulate quantum properties of solids, such as the valley degree of freedom, with ultrashort controlled lightwaves. Here we employ a sequence of few-optical-cycle visible pulses controlled with attosecond precision to excite and switch the valley pseudospin in a 2D semiconductor. We show that a pair of pulses separated in time with linear orthogonal polarizations can induce a valley-selective population. Additionally, exploiting a four-pump excitation protocol, we perform logic operations such as valley de-excitation and re-excitation at room temperature at rates as high as ~10 THz.

# Introduction

Decreasing the size of transistors has so far enabled steady increase in the speed and efficiency of information processing following Moore's law, but continuing to do so will soon become impractical: signal leaks, overheating, or malfunctions due to quantum effects, are some of the grim prospects of ultra-small electronic devices (*1*). One of the overarching possibilities that arise at the interface between ultrafast laser technology and 2D material science is the opportunity to realize light-driven switches of quantum properties (*2, 3*). Given that modern optical technologies allow one to control individual oscillations of the lightwave with sub-femtosecond precision, such light-controlled switches hold the potential to operate at nearly petahertz frequencies, orders of magnitude faster than the current standard, and on timescales shorter than electronic dephasing times at room temperature.

In this context, research on 2D materials has opened the possibility to use additional degrees of freedom, besides the electronic charge, which can provide increased information transport and storage capacity in the same space (*4-6*). In monolayer transition metal dichalcogenides (TMDs), the conduction and the valence bands display two energy-degenerate valleys which are located at crystal momenta $K$ and $K'$ at the corners of the Brillouin zone (*7-9*). The capability to selectively localize the particle in these regions of reciprocal space allows one to define a new binary index of the quantum state of the particle, the valley degree of freedom. Importantly, the three-fold rotational symmetry of the Bloch wavefunctions at $K$ and $K'$ gives rise to optical valley selection rules: light of $\sigma^+$ helicity couples only to the $K$ valley, while $\sigma^-$ light couples only to the $K'$ valley (*10*).

In addition, 2D TMDs exhibit enhanced Coulomb interactions due to their reduced dimensionality (*11*). This allows electron-hole pairs to bind together, forming bound quasiparticles known as excitons, which inherit from the charge carriers both the valley degree of freedom and the optical selection rules (*12-14*). Once initialized, the exciton must remain localized in the valley long enough to perform a function. A paradigmatic example of such a function would be a switch, where the valley polarization is turned on and off (*15-17*). Another relevant example is an amplifier, which increases the magnitude of the input signal. Both operations are ubiquitous in modern electronics, but they have proven to be exceptionally difficult to achieve for the valley degree of freedom. The main culprit is the short valley lifetime. Excitons remain localized in a specific valley for as little as 200 fs at room temperature before undergoing scattering to other crystal momenta (*18, 19*), mainly driven by exchange interactions (*20, 21*). Several routes have been explored to address this critical challenge: encapsulating TMD monolayers with wide-gap insulators to reduce environment interactions, engineering heterostructures that support longer-lived interlayer excitons, or using external fields (*6*). Recently, Langer et al. (*3*) demonstrated ultrafast population transfer between the valleys by means of a strong terahertz pulse. The laser-dressed state created in this way, however, is not well localized, but deformed and spread over many crystal momenta. Moreover, precise control over the valley switching in this configuration requires fine tuning of the frequency and intensity of the strong terahertz field, a challenge that is further exacerbated by the presence of non-linear effects.

Here, we use a sequence of weak, phase-locked, few-cycle, linearly polarized pulses with controlled time delays and polarizations to demonstrate a room-temperature valley switch and amplifier operating in less than 100 fs in a monolayer of the TMD WS$_2$. By controlling the delay between the pulses with sub-fs precision we selectively excite $K$ or $K'$, demonstrating that our method remains effective even when the pulses are fully temporally separated. All these operations are performed faster than the intervalley scattering and the excitonic dephasing times, both of which are measured independently in the time domain by our experimental protocol. Our work overcomes two critical hurdles for the implementation of practical valleytronic devices at room temperature: all-optical control of the valley degree of freedom and implementation of cascaded logic operations at a rate of tens of THz.

## Results

We investigate and manipulate the excitonic dynamics in a monolayer of WS$_2$, an atomically thin semiconductor characterized by a direct bandgap at the corners of the hexagonal Brillouin zone, known as $K$ and $K'$. Despite its complexity, the key physics of light interaction with the system can be described with a simple three-level model, schematically depicted in Fig. 1A. The ground state $|g\rangle$ represents the configuration where the valence band is full and the conduction band is empty, while the two excited degenerate states of energy $\hbar\omega_f$ represent the A-excitons at the $|K\rangle$ and $|K'\rangle$ valleys. A linearly-polarized pulse $E_x(t)$, see Fig. 1B, which is equivalent to a sum of left and right circularly polarized pulses, creates a coherent superposition of carriers in the $|K\rangle$ and $|K'\rangle$ states. The resulting state $|X\rangle$ can be schematically drawn as a pseudospin $\vec{s}$ lying on the equator of a Bloch sphere, as shown in Fig. 1D. This pseudospin represents the 2x2 exciton block $\rho_{K,K'}$ (dashed box in Fig. 1D) of the overall 3x3 density matrix.

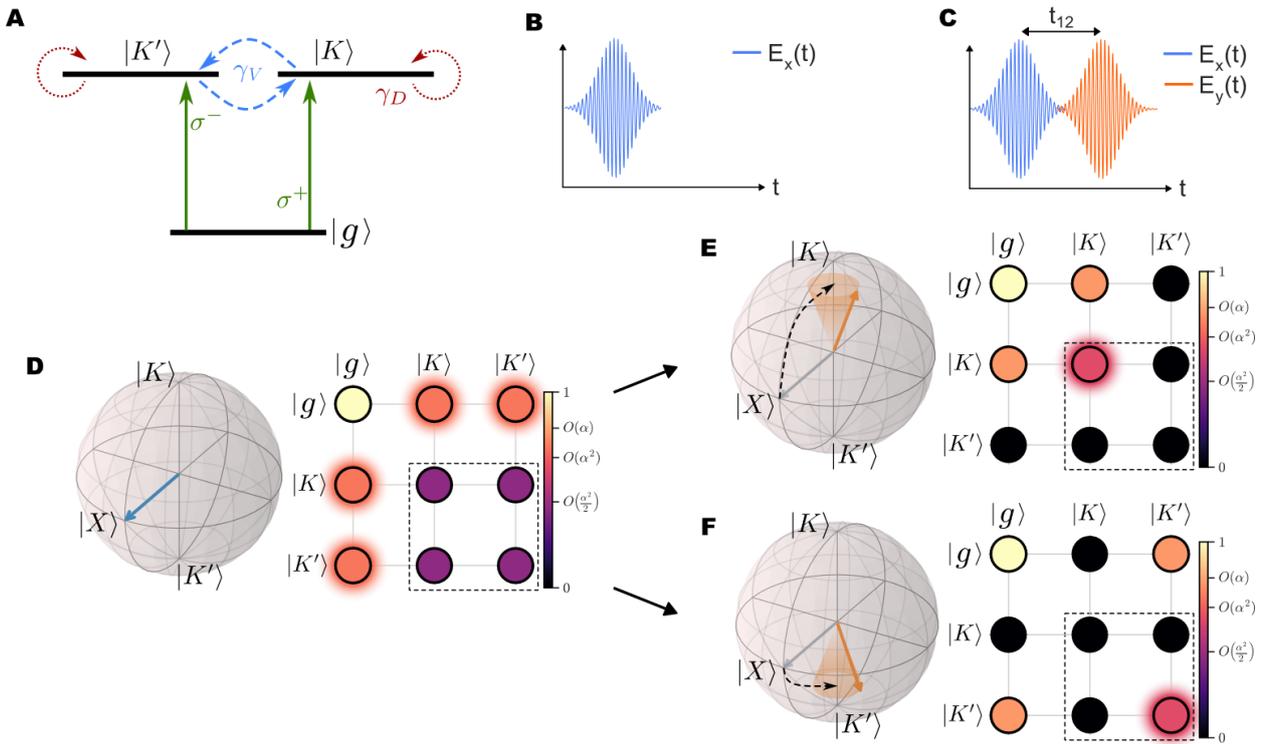

**Fig. 1. Theoretical model for all-optical valley control and two-pulse interaction.** (**A**) 3-level V-type system. The green arrows show the valley-selective transition in TMD monolayers. $\sigma^-$ and $\sigma^+$ couple respectively with the $K'$ and $K$ valleys. The dashed blue arrows depict the intervalley scattering, quantified by the rate $\gamma_V$, while $\gamma_D$ (red dashed arrows) accounts for exciton dephasing processes. The relaxation rate to the ground state is much smaller and is neglected. (**B** and **C**) Representation of the electric field profiles of the single (B) and double (C) pump pulse. (**D**, **E** and **F**) Bloch sphere and density matrix after the interaction with one (D) and two (E and F) pulses. The state $|X\rangle$, excited by a single linear pulse $E_x(t)$, shows an equal excitonic population $\alpha^2/2$ between the valleys, where $\alpha^2$ is the total excitation probability. The $K, K'$ block of the density matrix (dashed box) is represented by a pseudospin residing on the equatorial plane of the Bloch sphere. The density matrix $\mathcal{O}(\alpha)$ terms related to the light-induced coherence between ground and excited states have been highlighted with orange shades. The second pulse leads to a valley-selective excitation resulting in a rotation of the pseudospin towards $|K\rangle$ (E) or $|K'\rangle$ (F), depending on the delay between the pulses. The dominant valley population terms, residing on the diagonal of the density matrices in (E) and (F), have been highlighted.

Importantly, the linearly-polarized pulse establishes coherence between the ground and the excited $|K\rangle$ and $|K'\rangle$ states (highlighted terms in Fig. 1D). This coherence allows one to rotate the valley pseudospin in a controlled way by using a second pulse, delayed by a time $t_{12}$ relative to the first (*15*). In particular, if the second pulse is linearly polarized perpendicular to the first pulse ($E_y(t)$), as in Fig. 1C, the valley pseudospin can be rotated either to the north ($|K\rangle$) or to the south ($|K'\rangle$) poles of the Bloch sphere by using a proper choice of the delay $t_{12}$ (Fig. 1, E and F). Indeed, the effect of the pair of perpendicular, linearly polarized pulses is equivalent to that of a circularly-polarized pulse if the time delay $t_{12}$ between the pulses is such that $t_{12}\omega_f = (\pm\pi/2 + 2\pi N)$ with integer $N$. The sign $\pm$ in front of $\pm\pi/2$ controls the direction of rotation.

Notably, this remains true regardless of the temporal overlap between the two pulses as long as the excitonic coherence generated by the first pulse is maintained by the system until the arrival of the second pulse. Thus, on the one hand, Fig. 1E and F simply reflect the optical valley selection rules. On the other hand, in this scheme the sub-optical-cycle change in the exact timing of the two pump pulses can switch the sign of the valley polarization. Despite the apparent simplicity of the scheme, maintaining the excitonic coherence until the second pulse arrives is challenging. Indeed, the times of exciton dephasing $T_2$ and valley polarization decay $\tau$ in TMDs are only a few hundred femtoseconds at $T \sim 100$ K and are below 100 fs at room temperature (*18, 19, 22*). Implementing the two-pulse protocol requires ultrashort pulses with controllable timing and polarizations.

*Coherent control of the valley population*

The phase-locked pair of pump pulses is generated by the Translating-Wedge Based Identical Pulses eNcoding System (TWINS) birefringent interferometer (*23*) (see Fig. 2A and Methods) and their temporal separation (i.e. $t_{12}$) is controlled with sub-10-attosecond precision, while the delay between the probe pulse and pump pairs is $t_{\text{pr}}$. The valley polarization dynamics are measured using degenerate time-resolved Faraday rotation (TRFR), a pump-probe technique extensively used to study spin/valley-dependent dynamics in semiconductors and TMDs (*19, 24*). Typically, the pump pulse is circularly polarized and usually tuned on resonance with the excitonic transition, while the probe is linearly polarized and its polarization rotation, proportional to the spin/valley unbalance, is detected by an optical bridge, consisting of a Wollaston prism followed by a balanced photodetector. Here, the single circularly polarized pump pulse is replaced by a pair of ultrashort, phase-locked, linearly-polarized pulses with perpendicular polarizations, generated by the TWINS interferometer (see the sketch in Fig. 2A).

The imbalanced distribution of excitons between the *K* and *K'* valleys, induced by this pump pair, gives rise to the helicity-selective Pauli blocking and distinct refractive indexes for right- and left-circularly polarized probe pulses. When the linearly polarized probe pulse crosses the sample, its right- and left-circularly polarized components experience different phase shifts, leading to the rotation of the polarization plane. Therefore, the sign of the TRFR signal depends on the valley population imbalance, while its relaxation dynamics reflects the inter-valley (i.e. *K* to *K'*) scattering process (*19*). In our experiment, the pump pulses have a duration of ~18 fs and spectrum resonant with the A exciton transition, while the TRFR signal is detected at the energy of $2.03 \pm 0.017$ eV.

Figure 2B shows the TRFR signal acquired by scanning the $t_{12}$ delay while keeping $t_{\text{pr}} = 50$ fs fixed. A schematic representation of the pump and probe pulses is depicted in the inset on the top right corner. The TRFR signal is zero (corresponding to an equal excitonic population of both the valleys) when the two pump replicas temporally overlap (i.e. $t_{12} = 0$ fs) and oscillates around zero with a ~2 fs period defined by the A-exciton excitation frequency $\omega_f$. As anticipated, the extrema of the oscillation amplitude correspond to a delay $t_{12} = (\pm\pi/2 + 2\pi N)/\omega_f$ for which the pulse pair produces an effective circularly polarized field of $\pm$ helicity and induces a net valley polarization in *K* or *K'*, respectively. By continuously tuning $t_{12}$ we can change the valley pseudospin on the Bloch sphere from the poles across the equator. Indeed, the induced valley polarization can be expressed as $\sim 4\alpha^2 \sin(\omega_f t_{12}) e^{-t_{12}/T_2}$, where $\alpha$ is the excitation amplitude and $T_2$ is the dephasing time (see Methods).

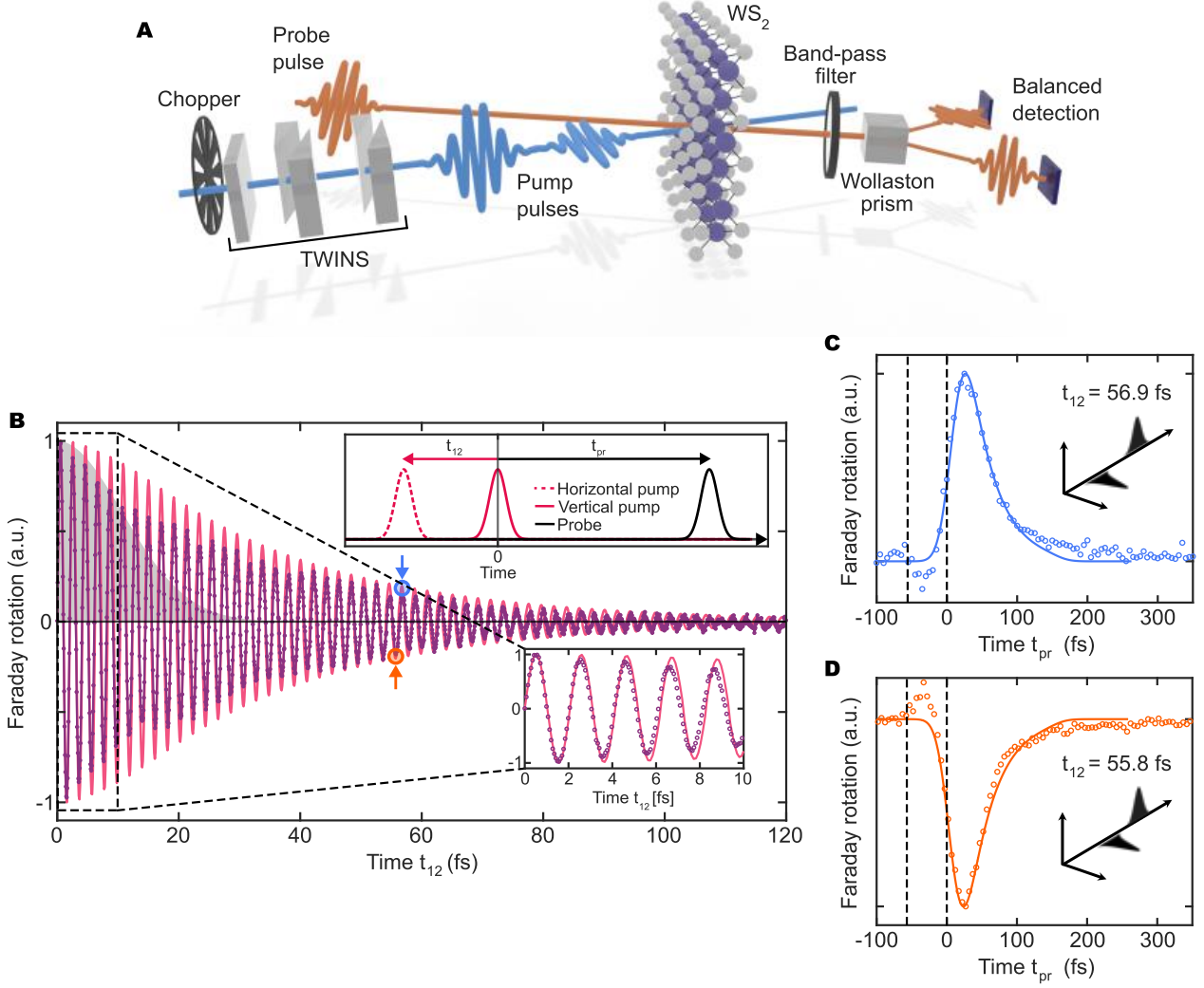

**Fig. 2. Control of valley polarization with phase-locked orthogonally polarized pulses**. Two-pulse time-resolved Faraday rotation experiment. (**A**) Double pump pulse generation scheme and TRFR setup. (**B**) TRFR experimental results obtained by continuously varying the delay between the pump pulses ($t_{12}$) and keeping fixed the probe pulse delay. The purple line connects the experimental points and acts as a guide for the eye. The pink solid line shows the results of the simulation (see Methods). The top-right panel shows the pulse sequence and relative delays. The inset in the bottom-right corner shows a zoom of the first 10 fs. The grey shaded area represents the estimated cross-correlation of the pump pulses. (**C** and **D**) TRFR data obtained by scanning $t_{pr}$ while keeping $t_{12}$ fixed and equal to 56.9 fs and 55.8 fs, corresponding respectively to a maximum and a minimum of the valley polarization (blue and orange arrows in panel (B)). The solid lines in the panels are simulated dynamics. The black dashed lines in the figure indicate the arrival times of the pump pulses.

We stress that this result is not a consequence of a mere temporal superposition of the two linearly polarized pulses, since the oscillations of the valley polarization persist well after their temporal overlap vanishes, as shown in Fig. 2B. This process results from the interaction between the electronic polarization induced by the first pump pulse and the second pump pulse. The decay of the oscillations is related to the dephasing of the electronic coherence and the decay of the valley polarization. The dots in Fig. 2C and D show the TRFR signals obtained by scanning the pump-probe delay ($t_{\text{pr}}$) with respect to the pump pair. In these measurements, the two pulses in the pair are delayed by a fixed $t_{12} = 56.9$ fs and $t_{12} = 55.8$ fs, corresponding respectively to a maximum and a minimum of the oscillations in Fig. 2B (highlighted by the blue and orange arrows).

These results are reproduced by the simulations of the three-level system depicted in Fig. 1A (solid lines on Fig. 2, C and D). We used the Lindblad formalism (see Methods) with two parameters, the valley polarization decay rate $\gamma_V = 1/(2\tau)$ and the pure dephasing rate $\gamma_D = 1/T_2^*$, where the total dephasing time is given by $1/T_2 = 1/(4\tau) + 1/T_2^*$. These two timescales $\tau$ and $T_2^*$ can be independently obtained from the experiment and govern the two critical dynamical processes. The first is the decay of the excited populations: as the excitons are selectively injected in a valley by a proper sequence of pump replicas, the valley polarization decays on a timescale of $\tau \approx 75$ fs. This value was extracted by comparison with the experimental results in Fig. 2, C and D. The fast exponential decay of the valley population is attributed to Maialle-Silva-Sham short-range exchange interaction (*20, 25*). The second process is related to the exciton polarization dephasing due to their scattering off the environment. The timescale for this process, $T_2^* \approx 34$ fs, is obtained by comparison with the experimental results in Fig. 2B. We emphasize that all experiments were performed at room temperature.

*Optical valley switch*

While the two-pulse setup demonstrates controllable initialization of the valley pseudospin, a fundamental step towards implementing ultrafast logical operations is the manipulation of the initialized valley state. We now demonstrate such ultrafast manipulation and reading with a four-pump-pulse and one-probe-pulse setup. In particular, we perform two fundamental operations with the valley degree of freedom: coherent switching and amplification. The protocol we adopt consists in manipulating the valley polarization through a sequence of four linearly-polarized pulses with the same intensity: two pairs of phase-locked pulses, each one characterized by mutually orthogonal polarizations (see Fig. 3A, magenta solid and dashed lines). Due to constraints related to the experimental setup, the temporal delay between the pulses is the same for each pair ($t_{12} = t_{34}$), while the delay between the two pairs ($t_{23}$) can be independently controlled (see Methods). All delays can be tuned with attosecond precision.

For clarity, in Fig. 3, B, C and D we show the results of the simulations for the four-pump-pulse protocol in two extreme regimes. In Fig. 3B, we show the trivial case where $t_{23}$ is much larger than the valley polarization decay and decoherence times. In this regime the valley polarization fully decays before the third pulse arrives and the final valley polarization does not depend on $t_{23}$. That is, the action of the two pump pairs is incoherent: the valley pseudospin can be initialized, but not manipulated. On the other hand, the solid lines in Fig. 3, C and D show the results of the fully coherent case, in which both the intervalley scattering and dephasing rates are negligible. In Fig. 3C, the third pulse acts as a $\pi$-pulse complement of the first pump pulse, rotating the valley pseudospin from the north pole at $|K\rangle$ to the equator along $|Y\rangle$, switching off the valley polarization. Due to the fact that the experimental setup imposes $t_{34} = t_{12}$, the fourth pulse also acts as $\pi$-pulse complement of the second pump pulse, effectively returning all of the valley population to the ground state $|g\rangle$ (coherent switching). In Fig. 3D, the pulse pair delay $t_{23}$ is such that the third and fourth pulses add coherently with the first and second pulses, resulting in the sum of the amplitudes of the two circular pulses (coherent amplification).

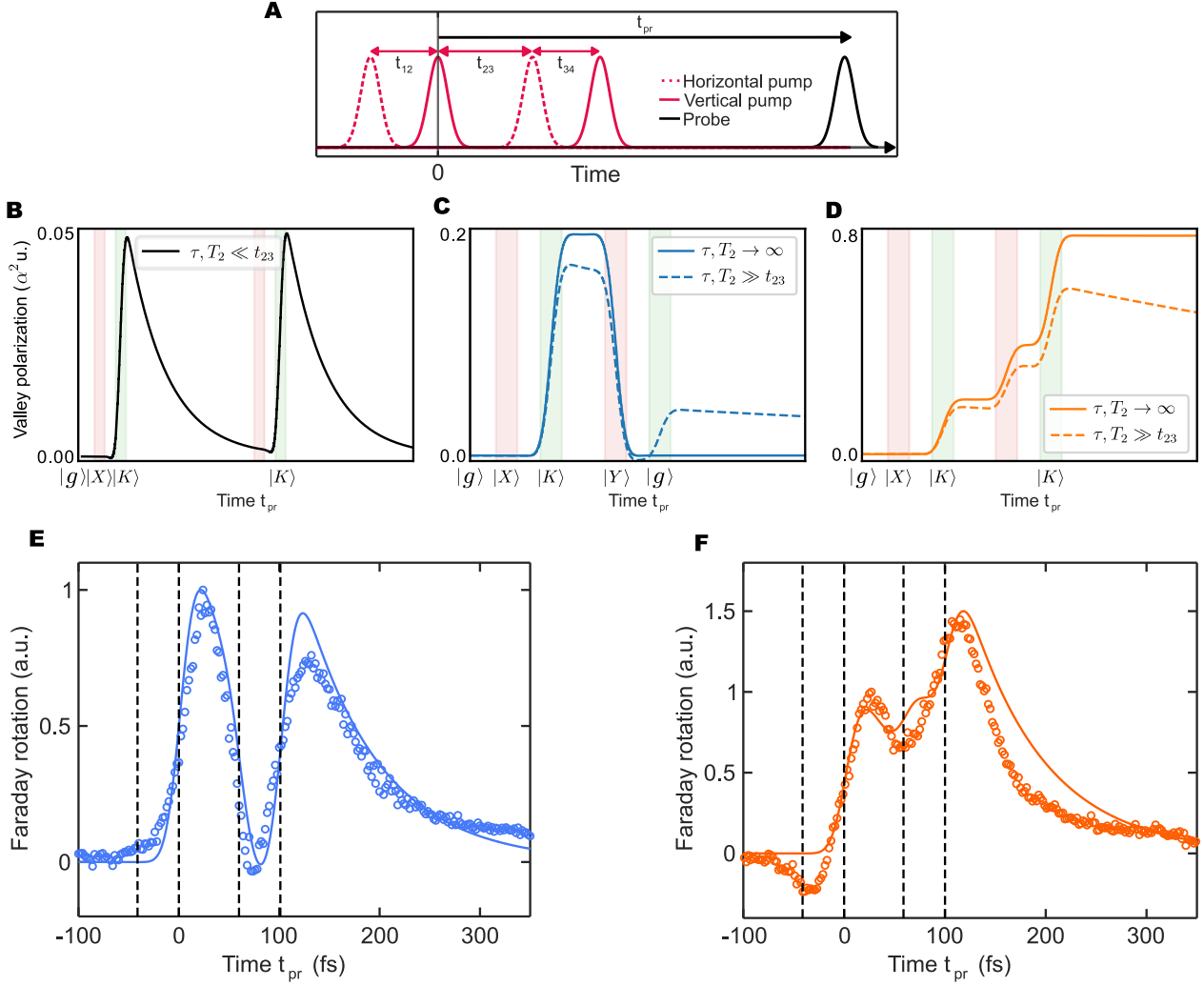

**Fig. 3. Coherent switch off and amplification of valley polarization.** Four-pulse TRFR experiments and simulations. (**A**) Pulse sequence used for the experiments and simulations. (**B**, **C** and **D**) Theoretical simulation of the four-pulse experiment in different scenarios. Panel (B) shows the incoherent case. The solid lines in panels (C, D) represent the fully coherent scenario, where $t_{23}$ is set to either switch off or amplify the valley polarization respectively. The dashed lines represent an intermediate case, where $t_{23}$ is still shorter than $\tau$ and $T_2$, but the intervalley scattering and the dephasing are not negligible. The results are obtained from the numerical solution of the 3-level model. The red/green shaded bands represent the FWHM of a pulse in the *x/y* direction. (**E** and **F**) The blue and orange dots illustrate the TRFR experiment obtained fixing the delays between the four pump pulses ($t_{12}$, $t_{23}$ and $t_{34}$) and continuously varying the pump-probe delay ($t_{pr}$). For both the panels, $t_{12} = t_{34} = 41.2$ fs. The arrival of the pump pulses is marked by the black dashed lines in the figures. The solid lines show the relative simulations. The two panels differ in $t_{23}$. In (E) $t_{23}$ is set equal to 59.9 fs, while in (F) $t_{23}$ is set to 58.9 fs.

Figures 3, E and F show the experimental results corresponding to the switching and amplification setups, respectively. Our experimental regime is clearly not the incoherent one displayed in Fig. 3B. Panel 3E shows a complete switch-off of the valley polarization in ~50 fs, while panel 3F shows a 50% amplification of the signal. The experimental protocol is not fully coherent, though, as evidenced by the fact that the valley polarization in Fig. 3E is restored after the fourth pulse, which does not occur for the fully coherent case in Fig. 3C (solid blue curve). As we show in dashed curves in Fig. 3, C and D, this scenario occurs when $t_{23}$ is smaller than the relaxation and dephasing times, but the effects of the latter are not negligible. In this case, the third pulse is able to switch off (panel 3C) or amplify (panel 3D) the valley polarization, but the gradual loss of coherence switches back on or caps

the amplification of the valley polarization (Fig. 3C and 3D, respectively), as observed in the experiment.

## Conclusions

We have demonstrated that a pair of phase-locked pulses with orthogonal linear polarizations separated in time can induce an excitonic imbalance between the $K$ and $K'$ valleys. By precisely controlling the delay between the pulses with sub-optical-cycle precision, it is possible to reverse the sign of the induced valley polarization, as measured by time-resolved Faraday rotation. Additionally, our experimental protocol enables the extraction of the excitonic decoherence time $T_2$ and the valley relaxation time $\tau$ independently. Furthermore, we have shown that using a third pulse phase-locked to the first two allows us to reset or maintain the valley polarization until the arrival of a fourth pulse, which can either switch it on again or amplify it. This method achieves a valley information processing rate exceeding 10 THz at room temperature.

By demonstrating all-optical control of the valley degree of freedom, our technique paves the way for the implementation of cascaded valley logic operations and realization of valleytronic devices. Moreover, this protocol can be extended to any broken-inversion-symmetric 2D semiconductor with a hexagonal honeycomb structure. By employing trains of attosecond, polarization-controlled pulses, it will be even possible to push the switching rates to the PHz range.

## Methods

*Experimental methods*

The valley polarization dynamics are studied via time-resolved Faraday rotation (TRFR) using the following setup. First, a regeneratively amplified Ti:sapphire laser generates ~100-fs pulses with 1-kHz repetition rate at 1.55 eV (800 nm) photon energy. These pulses are used to drive a non-collinear optical parametric amplifier (NOPA) which produces linearly polarized tunable pulses with a broadband spectrum that covers most of the visible range. We select a spectrum ranging from ~1.77 eV (700 nm) to ~2.29 eV (540 nm), to avoid the direct excitation of the B exciton, centered around 2.4 eV (517nm) (*7, 8*). The NOPA pulses are compressed to nearly transform-limited duration by a pair of chirped mirrors. The beam is then split into pump and probe pulses using a beam splitter.

The probe pulse goes through a computer-controlled delay line (Phisik Instrumente M-605.2DD), scanning the temporal range of interest. The pump pulse passes through a birefringent interferometer, the Translating-Wedge Based Identical Pulses eNcoding System (TWINS) (*23*). This device relies on a set of four wedges and a plate made of $\alpha$-barium-borate ($\alpha$-BBO). $\alpha$-BBO is a uniaxial birefringent crystal, therefore, it is characterized by an extraordinary refractive index along a specific axis, which differs from the ordinary one. When a linearly polarized pulse passes through a $\alpha$-BBO plate, its projections on the extraordinary and ordinary axes travel with different velocities. Therefore, by selecting an initial polarization at 45 degrees with respect to the $\alpha$-BBO axis, one can generate a couple of delayed perpendicularly polarized pulses. A second pair of $\alpha$-BBO wedges, cut with the optical axis along the beam propagation direction, ensures that the delay between the horizontally and vertically polarized pulses can be varied, while keeping the arrival time of the latter fixed.

Overall, the TWINS enables the generation of two phase-locked perpendicularly polarized replicas of the input pulse. In the standard operation of TWINS, these replicas are made to interfere by projecting them on the same polarization by a linear polarizer; here, the polarizer is removed to keep the polarizations orthogonal. The two replicas are interferometrically stable and the delay between them can be controlled on an attosecond scale. A chopper on the pump beam path reduces the repetition rate by half, allowing pump-probe measurements. A second pair of chirped mirrors is employed to compensate for the dispersion introduced by the TWINS, reaching a temporal full width at half maximum (FWHM) of the pulses of ~18 fs.

Pump and probe pulses are focused on the sample via spherical mirrors. Following the interaction, a bandpass filter (BPF) is employed to detect energies slightly off-resonance with respect to the optical gap of the semiconductor, where the Faraday rotation signal reaches its maximum (*26*). In particular, we employ a $610 \pm 5$ nm BPF, corresponding to $2.03 \pm 0.017$ eV. Afterwards, a Wollaston prism (WP) spatially separates the orthogonal polarization components of the incoming pulse with respect to its optical axis. The pulses are then focused on a pair of balanced photodiodes. For TRFR, the WP is rotated to ensure equal intensity on the photodiodes, in order to maximize the signal-to-noise ratio. The output signal is the difference between the signals measured by the two photodiodes, which is directly proportional to the rotation angle of the probe polarization. Finally, a lock-in amplifier demodulates the signal at 500 Hz, detecting the pump-induced changes in the probe polarization.

To generate a train of four pulses, we employ an extra birefringent plate (a $\alpha$-BBO crystal) preceded by a polarizer (set at 45° with respect to the $\alpha$-BBO optical axis) in the pump beam path, before the TWINS. In this way, we generate two perpendicularly polarized, phase-locked replicas of the incoming pulse, separated in time by ~100 fs, which corresponds to the delays between the first-third ($t_{13}$) and second-fourth ($t_{24}$) pulses in the final configuration. The time delay between the replicas can be slightly changed (from ~90 fs to ~110 fs) by rotating the $\alpha$-BBO crystal around its optical axis. The two replicas pass through another polarizer at 45°, giving rise to two collinear identical pulses. These pulses then enter the TWINS, generating two couples of orthogonally polarized pulses, separated by the same time delay ($t_{12} = t_{34}$).

The final result is a train of four pulses with horizontal and vertical polarization alternatively, where we can finely tune $t_{12}$ (equal to $t_{34}$) through the TWINS, and $t_{23}$ through the rotation of the birefringent crystal, hence varying its thickness. To precisely measure $t_{23}$ we record an interferogram of the four pulses using the TWINS. The first and third pulses are continuously shifted in time, precisely at negative delays, while the second and the fourth pulse remain fixed. Performing a measurement with a spectrometer we obtain an interferogram with two distinct peaks in amplitude. The first appears when $t_{12} = t_{34} = 0$ fs, thus the first and second (third and fourth) pulses are overlapped. The second arises when the second and third pulses are overlapped. The time distance between the two peaks in the interferogram corresponds to $t_{13}$ (which is equal to $t_{24}$). Since we know $t_{12}$ and $t_{34}$ with very high precision, we can obtain $t_{23}$ with a simple subtraction. The threshold for the linear excitation regime has been characterized through pump-probe measurements. We found a value equal to 30 µJ/cm$^2$. Throughout the whole experiment, the fluence of the pulses is kept below 10 µJ/cm$^2$.

*Pulses characterization*

To characterize the temporal duration of the pulses, we employ polarization-gated frequency resolved optical gating (PG-FROG). An additional polarizer at 45° was inserted in the pump beam path after the TWINS and $t_{12}$ was set to zero, thus obtaining a single pump pulse, while the probe polarization was set horizontal. The pulses were focused on a 160 µm thick fused silica window. The pump pulse induces a transient birefringence, increasing the refractive index only in the direction parallel to its polarization axis. This third order nonlinear effect causes a rotation of the probe polarization only when the pump and probe pulses overlap in time. A vertical polarizer, used as an analyzer, was inserted in the probe path before the spectrometer. Only when the probe polarization was rotated it could pass through the analyzer. The differential intensity, which is the probe intensity with pump minus the one without it, was measured to estimate the pulse duration. We obtained a full width at half maximum (FWHM) of the intensity profile of the pulses of ~17.8 fs, resulting in a cross-correlation FWHM of ~25.2 fs.

*Theoretical simulations*

Although the system demonstrates a relatively high level of complexity, its excitation dynamics can be fully represented by a three-level model. It has been established that the interaction between an exciton state in a semiconductor and light can be accurately explained using a two-level system (*27*). However, due to the inherent valley degree of freedom present in 2D materials with hexagonal symmetry (*28*), it is necessary to extend the model to incorporate this effect. Hence, instead of considering a single exciton, it becomes essential to account for two degenerate excitons (*18*). This configuration is often referred in the literature as the three-level *V* system (*29*), while its counterpart is known as the Λ system, as described in (*30*).

More formally, we consider the following system: a ground state, $|g\rangle$, which represents the electrons on the valence band, and two excited and degenerate states, $|K\rangle$ and $|K'\rangle$, which represent the excitonic states at each valley, respectively. Without loss of generality, we set the energy of $|g\rangle$ at 0 and the energy of $|K\rangle$ and $|K'\rangle$ at $\hbar\omega_f$. Thus, the Hamiltonian is simply:

$$H_0 = \hbar\omega_f(|K\rangle\langle K| + |K'\rangle\langle K'|) \tag{1}$$

As stated before, both valleys have different optical selection rules in the real material, so this must be represented within the model. Accordingly, the coupling with light must be:

$$H_L(t) = d\mathbf{E}(t) \cdot (|g\rangle\langle K| + |g\rangle\langle K'|, i|g\rangle\langle K| - i|g\rangle\langle K'|) + h.c., \tag{2}$$

where $d$ is the dipole coupling strength, $\mathbf{E}(t)$ is the electric field and $h.c.$ stands for hermitian conjugate. Hence, only light exhibiting right-handed circular polarization, specifically $\mathbf{e}_1 + i\mathbf{e}_2$, will interact with the $|K\rangle$ valley. Conversely, if a left-handed circularly polarized laser field is used, it will couple with the $|K'\rangle$ valley. We have set the dipole strength to $d = 1$.

The dynamics will be solved using a Lindblad master equation (*31*) of form:

$$\frac{d\rho}{dt} = -i[H(t), \rho(t)] + \sum_n L_n\rho(t)L_n^\dagger - \frac{1}{2}\{L_n^\dagger L_n, \rho(t)\}, \tag{3}$$

where $\rho(t)$ is the density matrix, $H(t) = H_0 + H_L(t)$ is the time-dependent hamiltonian of the system and $L_n$ are the collapse operators. Our initial condition will be $\rho(0) = |g\rangle\langle g|$. The observable that we will pay most attention to is the imbalance of the population between both valleys. More precisely,

$$\sigma(t) = \text{Tr}\{P_K\rho(t)\} - \text{Tr}\{P_{K'}\rho(t)\}, \tag{4}$$

where $P_{K/K'}$ are the projectors onto the excited states.

The selection of collapse operators $L_n$ is crucial, as they will represent intervalley scattering and dephasing, two significant physical phenomena in this system. To properly include both phenomena, four distinct operators will be required. We will start by considering the operators related to intervalley scattering. Those operators have the following form (*32*)

$$R_{ml} = \sqrt{\gamma_{ml}} |m\rangle\langle l|. \tag{5}$$

In our case, we will have two of those operators, namely

$$\begin{cases} R_{K'K} = \sqrt{\gamma_V} |K'\rangle\langle K| \\ R_{KK'} = \sqrt{\gamma_V} |K\rangle\langle K'| \end{cases} \tag{6}$$

Their physical interpretation can be easily understood if one applies any of those operators to an excited state: $R_{K'K}|K\rangle = \sqrt{\gamma_V}|K'\rangle$, i.e., they prompt the system to jump to the opposite valley with a rate of $\gamma_V$. This rate $\gamma_V$ remains equal for both states, given their energy degeneracy. The parameter $\gamma_V$ is related to the valley polarization decay time, $\tau$, as $\gamma_V = 1/(2\tau)$.

To include pure dephasing (as intervalley scattering will always include some factor of dephasing by itself), one needs to take into account operators of the form (*32*)

$$D_n = \sqrt{\gamma_m} |m\rangle\langle m|. \tag{7}$$

For our system, they will be:

$$\begin{cases} D_g = \sqrt{\gamma_D} |g\rangle\langle g| \\ D_K = \sqrt{\gamma_D} |K\rangle\langle K| \\ D_{K'} = \sqrt{\gamma_D} |K'\rangle\langle K'| \end{cases} \tag{8}$$

The physical interpretation becomes clearer when one plugs one of these operators into the collapse term of Eq. 3. This action reveals that the $D_n$ operators eliminate the off-diagonal terms, which correspond precisely to the coherence terms (*31*). The parameter $\gamma_D$ is associated with the pure valley dephasing time, $T_2^*$, through the relationship $\gamma_D = 1/T_2^*$, with the total dephasing time $T_2$ given by $1/T_2 = \frac{1}{2}\gamma_V + \gamma_D = 1/(4\tau) + 1/T_2^*$.

The physical parameters of the model, namely $\hbar\omega_f$, $\tau$, and $T_2$, were determined through fitting to experimental data of the two-pulse experiment. The relaxation time, $\tau$, was obtained when the two fields were spatially overlapped, precisely at $t_{12} = 0.5$ fs. In this configuration, the total field becomes completely circular. Therefore, the valley polarization will decay after the pulse following an exponential law given by $e^{-t/\tau}$. The value obtained was $\tau \approx 75.16$ fs. Lastly, we fitted the scan of maximum of the Faraday rotation in terms of $t_{12}$. Such maximum must should oscillate as

$$\sigma_{\max}(t_{12}) \propto \sin(\omega_f t_{12}) e^{-t_{12}/T_2} \tag{9}$$

The values obtained from that fit were: $T_2^* \approx 33.51$ fs and $\hbar\omega_f \approx 1.98$ eV. The fitted binding energy of the exciton is in accordance to the one measured in the experiment, $\hbar\omega_f^{(\text{expt})} \approx 2.01$ eV, using the linear absorption spectrum.

The laser parameters were chosen to match the experimental ones. We chose a Gaussian envelope with a FWHM of 20 fs and an intensity such that we are in the linear regime. The central laser frequency was chosen to be $\hbar\omega_L = 2.03$ eV. Furthermore, to simulate the effect of the finite temporal resolution of the system, the theoretical results have been convoluted with Gaussian envelope with a FWHM of 24 fs.

**Acknowledgments:** R.E.F.S. and P.S-J. acknowledge support from Grants PID2021-122769NB-I00 and RYC-2022-035373-I, funded by MICIU/AEI/10.13039/501100011033, 'ERDF A way of making Europe' and 'ESF+'. A.J.-G. acknowledges support from the Talento Comunidad de Madrid Fellowship 2022-T1/IND-24102 and the Spanish Ministry of Science, Innovation and Universities through grant reference PID2023-146676NA-I00. G.C. and S.D.C. acknowledge support from the European Union's NextGenerationEU Programme with the I-PHOQS Infrastructure [IR0000016, ID D2B8D520, CUP B53C22001750006] "Integrated infrastructure initiative in Photonic and Quantum Sciences". G.C. acknowledges support by the acknowledge the Horizon Europe European Innovation Council (101130384, HORIZONEIC-2023-PATHFINDEROPEN-01, QUONDENSATE). S.D.C. acknowledges support from the European Union's NextGenerationEU – Investment 1.1, M4C2 - Project n. 2022LA3TJ8 – CUP D53D23002280006.

**Author contributions:** M.I. and G.C. conceived and supervised the project. F.G. performed the experiments with the contribution of M.R. and F.V.A.C., under the supervision of S.D.C.. E.B.M. performed the numerical simulations with the theoretical support of P.S.-J. under the supervision of A.J.-G. and R.E.F.S.. All authors contributed to the discussion of the results and the writing of the paper.

**Competing interests:** Authors declare that they have no competing interests.

**Data availability:** The data that support the plots within this paper and other findings of this study are available from the corresponding authors upon reasonable request.

# References

1. M. M. Waldrop, The chips are down for Moore's law. *Nature News* **530**, 144 (2016).

2. Z. Ye, D. Sun, T. F. Heinz, Optical manipulation of valley pseudospin. *Nature Physics* **13**, 26–29 (2017).

3. F. Langer, C. P. Schmid, S. Schlauderer, M. Gmitra, J. Fabian, P. Nagler, C. Schüller, T. Korn, P. G. Hawkins, J. T. Steiner, U. Huttner, S. W. Koch, M. Kira, R. Huber, Lightwave valleytronics in a monolayer of tungsten diselenide. *Nature* **557**, 76–80 (2018).

4. I. Žutić, J. Fabian, S. D. Sarma, Spintronics: Fundamentals and applications. *Reviews of Modern Physics* **76**, 323 (2004).

5. J. R. Schaibley, H. Yu, G. Clark, P. Rivera, J. S. Ross, K. L. Seyler, W. Yao, X. Xu, Valleytronics in 2D materials. *Nature Reviews Materials* **1**, 1–15 (2016).

6. S. A. Vitale, D. Nezich, J. O. Varghese, P. Kim, N. Gedik, P. Jarillo-Herrero, D. Xiao, M. Rothschild, Valleytronics: opportunities, challenges, and paths forward. *Small* **14**, 1801483 (2018)

7. K. F. Mak, C. Lee, J. Hone, J. Shan, T. F. Heinz, Atomically thin MoS$_2$: a new direct-gap semiconductor, *Physical Review Letters* **105**, 136805 (2010).

8. A. Splendiani, L. Sun, Y. Zhang, T. Li, J. Kim, C. Chim, G. Galli, F. Wang, Emerging photoluminescence in monolayer MoS$_2$. *Nano Letters* **10**, 1271–1275 (2010).

9. A. Kormányos, G. Burkard, M. Gmitra, J. Fabian, V. Zólyomi, N. D. Drummond, V. Fal'ko, k·p theory for two-dimensional transition metal dichalcogenide semiconductors. *2D Materials* **2**, 022001 (2015)

10. D. Xiao, M.-C. Chang, Q. Niu, Berry phase effects on electronic properties. *Reviews of Modern Physics* **82**, 1959 (2010).

11. A. Chernikov, T. C. Berkelbach, H. M. Hill, A. Rigosi, Y. Li, B. Aslan, D. R. Reichman, M. S. Hybertsen, T. F. Heinz, Exciton binding energy and nonhydrogenic Rydberg series in monolayer WS$_2$. *Physical Review Letters* **113**, 076802 (2014).

12. H. Zeng, J. Dai, W. Yao, D. Xiao, X. Cui, Valley polarization in MoS$_2$ monolayers by optical pumping. *Nature Nanotechnology* **7**, 490–493 (2012).

13. K. F. Mak, K. He, J. Shan, T. F. Heinz, Control of valley polarization in monolayer MoS$_2$ by optical helicity. *Nature Nanotechnology* **7**, 494–498 (2012).

14. T. Cao, G. Wang, W. Han, H. Ye, C. Zhu, J. Shi, Q. Niu, P. Tan, E. Wang, B. Liu, J. Feng, Valley-selective circular dichroism of monolayer molybdenum disulphide. *Nature Communications* **3**, 887 (2012

15. R. E. Silva, M. Ivanov, Á. Jiménez-Galán, All-optical valley switch and clock of electronic dephasing. *Optics Express* **30**, 30347–30355 (2022).

16. N. Rana, G. Dixit, All-optical ultrafast valley switching in two-dimensional materials. *Physical Review Applied* **19**, 034056 (2023).

17. S. Sharma, P. Elliott, S. Shallcross, Valley control by linearly polarized laser pulses: example of WSe$_2$. *Optica* **9**, 947–952 (2022).

18. K. Hao, G. Moody, F. Wu, C. K. Dass, L. Xu, C.-H. Chen, L. Sun, M.-Y. Li, L.-J. Li, A. H. MacDonald, X. Li, Direct measurement of exciton valley coherence in monolayer WSe$_2$. *Nature Physics* **12**, 677–682 (2016)

19. S. Dal Conte, F. Bottegoni, E. A. A. Pogna, D. De Fazio, S. Ambrogio, I. Bargigia, C. D'Andrea, A. Lombardo, M. Bruna, F. Ciccacci, A. C. Ferrari, G. Cerullo, M. Finazzi, Ultrafast valley relaxation dynamics in monolayer MoS$_2$ probed by nonequilibrium optical techniques. *Physical Review B* **92**, 235425 (2015).

20. M. Z. Maialle, E. A. de Andrada e Silva, L. J. Sham, Exciton spin dynamics in quantum wells. *Physical Review B* **47**, 15776 (1993).

21. T. Yu, M. Wu, Valley depolarization due to intervalley and intravalley electron-hole exchange interactions in monolayer MoS$_2$. *Physical Review B* **89**, 205303 (2014)

22. M. Selig, G. Berghäuser, A. Raja, P. Nagler, C. Schüller, T. F. Heinz, T. Korn, A. Chernikov, E. Malic, A. Knorr, Excitonic linewidth and coherence lifetime in monolayer transition metal dichalcogenides. *Nature Communications* **7**, 13279 (2016).

23. D. Brida, C. Manzoni, G. Cerullo, Phase-locked pulses for two-dimensional spectroscopy by a birefringent delay line. *Optics Letters* **37**, 3027–3029 (2012).

24. S. A. Bourelle, F. V. A. Camargo, S. Ghosh, T. Neumann, T. W. J. van de Goor, R. Shivanna, T. Winkler, G. Cerullo, F. Deschler, Optical control of exciton spin dynamics in layered metal halide perovskites via polaronic state formation. *Nature Communications* **13**, 3320 (2022)

25. H. Yu, G.-B. Liu, P. Gong, X. Xu, W. Yao, Dirac cones and Dirac saddle points of bright excitons in monolayer transition metal dichalcogenides. *Nature Communications* **5**, 3876 (2014).

26. G. Plechinger, P. Nagler, A. Arora, R. Schmidt, A. Chernikov, A. G. del Águila, P. C. M. Christianen, R. Bratschitsch, C. Schüller, T. Korn, Trion fine structure and coupled spin–valley dynamics in monolayer tungsten disulfide. *Nature Communications* **7**, 12715 (2016).

27. M. Kira, S. W. Koch, *Semiconductor Quantum Optics* (Cambridge University Press, 2011)

28. X. Xu, W. Yao, D. Xiao, T. F. Heinz, Spin and pseudospins in layered transition metal dichalcogenides. *Nature Physics* **10**, 343–350 (2014).

29. K. Rzaźewski, Atoms and laser light: The theory of coherent atomic excitation. *Science* **250**, 1603–1603 (1990)

30. A. Vivas-Viaña, A. González-Tudela, C. S. Muñoz Unconventional mechanism of virtual-state population through dissipation. *Physical Review A* **106**, 012217 (2022).

31. H.-P. Breuer, F. Petruccione, *The Theory of Open Quantum Systems* (Oxford University PressOxford, 2007)

32. D. G. Tempel, A. Aspuru-Guzik, Relaxation and dephasing in open quantum systems time-dependent density functional theory: Properties of exact functionals from an exactly-solvable model system. *Chemical Physics* **391**, 130–142 (2011).